\documentclass[a4paper,superscriptaddress,twocolumn,prl,noshowkeys]{revtex4}
\pdfoutput=1
\usepackage{graphicx}
\usepackage{booktabs}
\usepackage{url}
\usepackage{float}
\usepackage{amssymb}
\usepackage{amsmath}
\usepackage{bm}
\usepackage{url}
\usepackage{setspace}

\begin{document}

\title{Few-Cycle Pulse Generation in an X-Ray Free-Electron Laser}

\author{D. J. Dunning}
\email{david.dunning@stfc.ac.uk}
\affiliation{ASTeC, STFC Daresbury Laboratory and Cockcroft Institute, Warrington, WA4 4AD, United Kingdom}
\affiliation{Department of Physics, SUPA, University of Strathclyde, Glasgow G4 0NG, United Kingdom}

\author{B. W. J. McNeil}
\email{b.w.j.mcneil@strath.ac.uk}
\affiliation{Department of Physics, SUPA, University of Strathclyde, Glasgow G4 0NG, United Kingdom}

\author{N. R. Thompson}
\email{neil.thompson@stfc.ac.uk}
\affiliation{ASTeC, STFC Daresbury Laboratory and Cockcroft Institute, Warrington, WA4 4AD, United Kingdom}
\affiliation{Department of Physics, SUPA, University of Strathclyde, Glasgow G4 0NG, United Kingdom}
\date{\today}

\begin{abstract}

A method is proposed to generate trains of few-cycle x-ray pulses from a Free-Electron Laser (FEL) amplifier via a compact `afterburner' extension consisting of several few-period undulator sections separated by electron chicane delays. Simulations show that in the hard x-ray (wavelength $\sim$0.1~nm; photon energy $\sim$10~keV) and with peak powers approaching normal FEL saturation (GW) levels, root mean square pulse durations of 700~zeptoseconds may be obtained. This is approximately two orders of magnitude shorter than that possible for normal FEL amplifier operation. The spectrum is discretely multichromatic with a bandwidth envelope increased by approximately two orders of magnitude over un-seeded FEL amplifier operation. Such a source would significantly enhance research opportunity in atomic dynamics, and push capability towards nuclear dynamics.

\end{abstract}
\maketitle

Pulses of light, tens to hundreds of attoseconds in duration, have enabled the exploration and control of processes that occur at atomic time scales~\cite{atto1,atto2}.
A common source of such pulses results from High Harmonic Generation (HHG) in a laser driven gas~\cite{hhg} from which isolated pulses may be generated or, more commonly, a periodic train of pulses which can act as an ultra-fast strobe.
This fast stroboscopic property has been successfully applied to a range of experiments to image and control electron wave packet behaviour in atoms~\cite{strobe, apts}. Reducing pulse durations towards 1 attosecond, and beyond into the zeptosecond regime with high (GW) peak-powers  may extend opportunities to directly resolve electronic behaviour within inner shells of atoms; the imaging and possible control of electronic-nuclear interactions such as Nuclear Excitation by Electron Transition/Capture (NEET/NEEC)~\cite{neet}; and move towards the resolution of nuclear dynamics~\cite{kaplan}. However, this will require a sufficient flux of photons with  energies in the hard x-ray ($\gtrsim$10~keV) which are not available from HHG sources.

The recently realised x-ray Free Electron Laser (FEL), which can generate the higher photon energies at multi-GW powers, would offer this enhanced temporal resolution if few-cycle, hard x-ray radiation pulses could be generated.
The FEL is currently a unique source for scientific experiments, with facilities such as FLASH~\cite{FLASH}, LCLS~\cite{LCLS}, SACLA~\cite{SACLA} and FERMI$@$elettra~\cite{FERMI@elettra} in operation, and others in development~\cite{natphoton}, including proposals for a very hard x-ray source of coherent 50~keV photons~\cite{marie}.
The normal x-ray FEL operating mode is via a high-gain amplifier generating Self-Amplified Spontaneous Emission (SASE)~\cite{bnp} which has noisy temporal and spectral properties~\cite{bonprl}, although new methods are being introduced to improve on this~\cite{natphoton}.
The characteristic minimum pulse duration for such high-gain amplifier FELs is determined by the FEL bandwidth~\cite{bonprl,limit}, which for present x-ray FELs corresponds to durations $\gtrsim 100$~as.
In this Letter a new operating method is proposed which, via a relatively simple upgrade, would allow existing x-ray FELs to generate trains of high-power (GW), few-cycle pulses into the zeptosecond regime - at least two orders of magnitude shorter than currently achievable.
The corresponding spectrum is discretely multichromatic within a broad bandwidth envelope.
Potential applications of such sources are the stroboscopic interrogation of matter~\cite{strobe} with intensities enhanced by orders of magnitude compared with current sources, while the multiple narrow frequency modes may be exploited in applications such as resonant inelastic x-ray scattering~\cite{RIXS}.
High energy photon pulses of zeptosecond duration begins to make feasible access to the temporal behaviour of the nucleus, in what has been coined nuclear quantum optics~\cite{nqo}.

In a high-gain FEL amplifier, a relativistic electron beam propagates through a long undulator, allowing a resonant, co-operative interaction with a co-propagating radiation field of resonant wavelength $\lambda_r=\lambda_u(1+\bar{a}_u^2)/2\gamma_0^2$~\cite{natphoton}, where $\lambda_u$ is the undulator period, $\bar{a}_u$ is the rms undulator parameter and $\gamma_0$ is the mean electron energy in units of the electron rest mass energy.
The co-operative instability results in an exponential amplification of both the resonant radiation intensity and the electron micro-bunching, $b=\langle e^{-i\theta_j}\rangle$~\cite{bnp}, where $\theta_j$ is the ponderomotive phase~\cite{natphoton} of the $j^\mathrm{th}$ electron.
In the one-dimensional limit, the length-scale of the exponential gain is determined by the gain length $l_g=\lambda_u/4\pi\rho$, where $\rho$ is the FEL coupling parameter~\cite{bnp}.
A resonant radiation wavefront propagates ahead of the electrons at a rate of $\lambda_r$ per $\lambda_u$. This relative propagation, or `slippage' in one gain length $l_{g}$ is called the `co-operation length', $l_{c}=\lambda_r/4\pi\rho$~\cite{bmp}, which determines the phase coherence length and bandwidth of the interaction.

\begin{figure*}[t]
   \centering
   \includegraphics[width=0.74\textwidth]{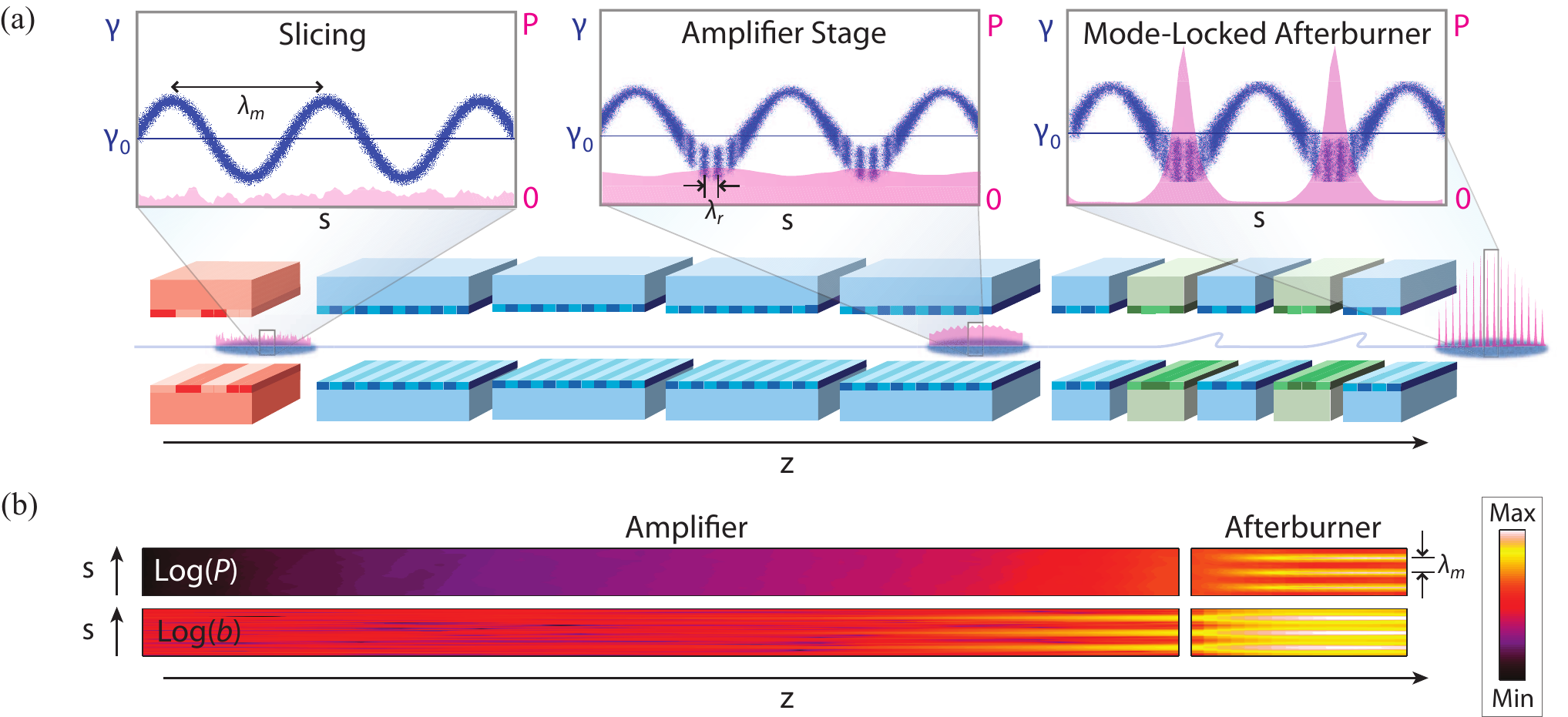}
   \vspace{-2mm}
   \caption{(a) Schematic layout of the proposed technique and (b) Example simulation results. An electron beam is sliced (e.g. using an external laser and a short undulator to apply an energy modulation, as shown), such that a comb structure develops in the FEL-induced electron micro-bunching ($b$) in a long undulator (amplifier stage). Further amplification of the radiation intensity ($P$) with periodic electron delays (mode-locked afterburner stage) generates a train of few-cycle radiation pulses.}
   \label{Schematic}
   \vspace{-5.5mm}
\end{figure*}

Several methods have been proposed to generate short radiation pulses by `slicing' short regions of high beam quality from within a longer electron pulse~\cite{saldin,zholentsNJP,emma,ml_esase}, with the shortest pulses generated at LCLS to date of $\gtrsim 1$fs duration~\cite{shortpulselcls}.
However, the FEL bandwidth restricts the minimum pulse length from such schemes to $\gtrsim l_{c}$~\cite{bonprl,limit}, with typical value $l_c \sim 200 \lambda_r$ corresponding to $\sim 100$~as for x-ray FELs.
By inserting electron delays between modules within the long undulator, the phase coherence length of the interaction can be discretised, increasing the bandwidth. The mode-locked FEL amplifier (ML-FEL) proposal~\cite{mlsase} uses this to generate a train of pulses with lengths $\ll l_{c}$ and peak powers approaching FEL saturation. The number of optical cycles per pulse is approximately the number of undulator periods per module, so could potentially deliver few-cycle pulses. However, this would require significantly modifying existing FELs, which typically have several hundred periods per module.

In this Letter, a method, shown schematically in Fig.~\ref{Schematic}, is proposed to generate trains of few-cycle radiation pulses similar to that of the ML-FEL but by using a short `afterburner' extension that could relatively easily be added to existing facilities.
The technique involves preparing an electron beam with periodic regions of high beam quality, each region of length $\ll l_{c}$, prior to injection into a normal FEL amplifier. Only these high quality regions undergo a strong FEL interaction within the amplifier to generate a periodic comb structure in the FEL-induced micro-bunching. Once the micro-bunching comb is sufficiently well developed, but before any saturation of the FEL process, the electron beam is injected into a `mode-locked afterburner', which maps the comb structure of the electron micro-bunching into a similar comb of the radiation intensity.
The afterburner comprises a series of few-period undulator modules separated by electron delay chicanes similar to that used in the ML-FEL~\cite{mlsase}.
These undulator-chicane modules maintain an overlap between the comb of bunching electrons and the developing radiation comb, each pulse of length $\ll l_c$, allowing it to grow exponentially in power to saturation.
The pulses are delivered in trains, since amplification occurs over a number of afterburner modules, and would be naturally synchronised to the modulating laser (Fig.~\ref{Schematic}).

Several methods could be used to generate the periodically bunched electron beam prior to the afterburner, including energy modulation~\cite{saldin, s2e_mlfel}, emittance spoiling~\cite{emma} and current enhancement~\cite{esase,njp_ml_curmod}.
Here, electron beam energy modulation is used as illustrated in Fig.~\ref{Schematic}.
For a sufficiently large sinusoidal energy modulation of the form $\gamma(t)=\gamma_0+\gamma_m\cos(\omega_mt)$, where $\omega_m$ is the modulation frequency, those regions of the beam about the mean energy $\gamma_0$ will be `spoiled' due to the larger energy gradients, whereas about the extrema, $\gamma\approx \gamma_0\pm\gamma_m$, a higher beam quality exists due to smaller energy gradients.
Only these latter regions may be expected to experience a strong FEL interaction within the amplifier to generate the comb structure in the electron bunching parameter.
In fact, strong micro-bunching develops only at the minima of the energy modulation. It is noted from FEL linear theory~\cite{bnp} that there is an asymmetry about the resonant frequency for the rate of exponential gain with a critical radiation frequency below which no exponential instability exists.
It may be intuitively expected that electrons about the minima will experience radiation fields generated by higher energy electrons, and so greater than their resonant frequency.
Due to this gain asymmetry favouring higher frequencies these lower energy regions of the modulated beam may be expected to dominate any FEL interaction in the amplifier.
This is what is observed in simulations here and in other work~\cite{s2e_mlfel} and has also been confirmed in a more complete linear theory for an energy modulated beam~\cite{linear}.

\begin{figure}[h!]
\vspace{4mm}
   \centering
   \includegraphics[width=0.92\columnwidth]{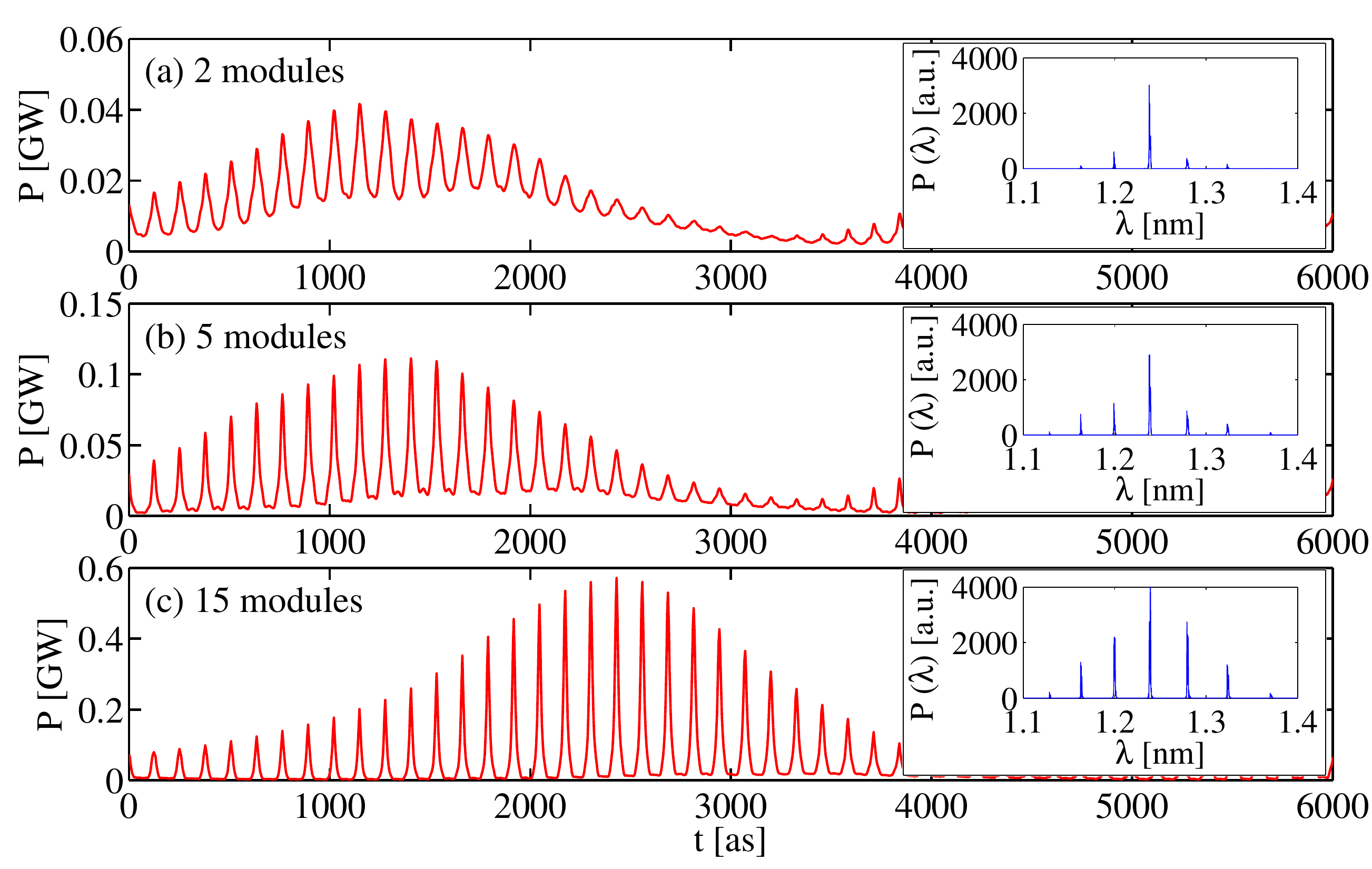}
   \vspace{-3mm}
   \caption{Soft x-ray mode-locked afterburner simulation results: Radiation power profile and spectrum after (a) 2, (b) 5 and (c) 15 undulator-chicane modules. The duration of an individual pulse after 15 modules is $\sim$9~as~rms.}
   \label{ML_AB_output}
\vspace{-7mm}
\end{figure}

Simulations of the method were carried out using the simulation code Genesis~1.3~\cite{genesis} using the parameters of Table~\ref{table1} in both the soft and hard x-ray. For the soft x-ray case, with resonant FEL wavelength of $\lambda_r$=1.24~nm, the modulation period $\lambda_m=38.44\mathrm{nm}~(=31\lambda_r)$ and $\lambda_m\ll l_c$.
\newcommand\T{\rule{0pt}{2.6ex}}
\newcommand\B{\rule[-1.2ex]{0pt}{0pt}}
\begin{table}[h!]
\small
\begin{spacing}{0.81}
\vspace{-3mm}
\caption{\label{table1}Parameters for soft and hard x-ray simulations.}
\begin{ruledtabular}
\begin{tabular}{l c c}
\textbf{Parameter}  \T \B & \textbf{Soft x-ray} & \textbf{Hard x-ray}\\
\hline
\emph{Amplifier stage}\T \B&\\
Electron beam energy [GeV] & 2.25 & 8.5\\
Peak current [kA] & 1.1 & 2.6\\
$\rho$-parameter &1.6$\times 10^{-3}$& 6$\times 10^{-4}$\\
Normalised emittance [mm-mrad] & 0.3 & 0.3\\
RMS energy spread, $\sigma_\gamma/\gamma_0 $ & 0.007~\%  & 0.006~\%\\
Undulator period, $\lambda_u$ [cm] & 3.2 & 1.8\\
Undulator periods per module & 78 & 277\\
Resonant wavelength, $\lambda_r$ [nm] & 1.24 & 0.1\\
Modulation period, $\lambda_m$ [nm] & 38.44 & 3\\
Modulation amplitude, $\gamma_m/\gamma_0$ & 0.1~\% & $0.06~\%$\\
Extraction point [m] & 34.1 & 36.0\\
\hline
\emph{Mode-locked afterburner}\T \B&\\
Undulator periods per module & 8 & 8\\
Chicane delays [nm] & 28.52 & 2.2\\
No. of undulator-chicane modules \B& $\sim$15 & $\sim$40\\
\end{tabular}
\end{ruledtabular}
\end{spacing}
\vspace{-5mm}
\end{table}
This modulation could be achieved using a modulating undulator seeded by current HHG sources~\cite{atto2,CDR}.
The undulator and electron beam parameters are those of the UK New Light Source design~\cite{CDR}. The electron pulse length was $\gg l_c$ with no other longitudinal variation of beam parameters.
\begin{figure}[t!]
   \centering
   \includegraphics[width=0.83\columnwidth]{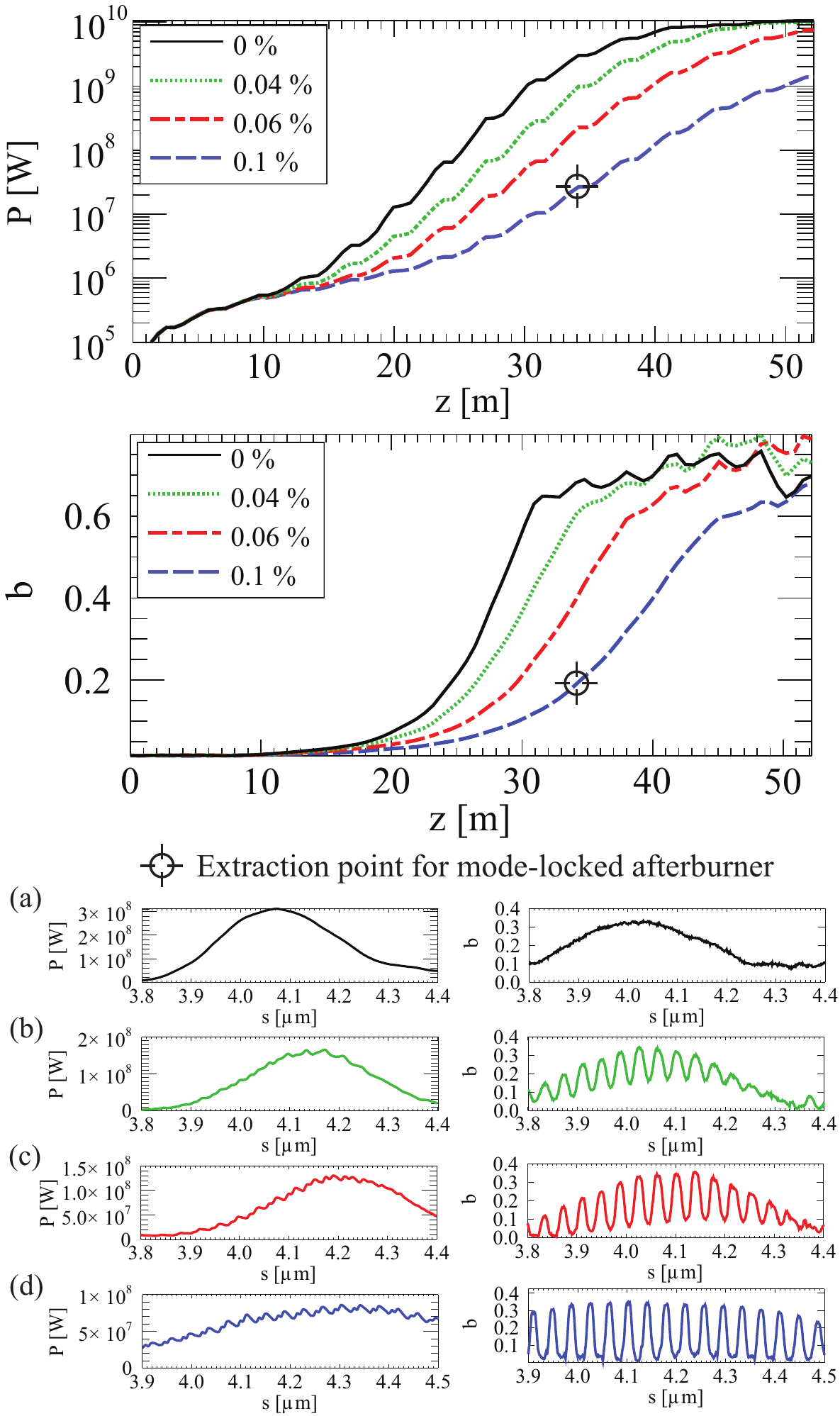}
   \vspace{-2mm}
   \caption{Optimisation of the amplifier stage for the soft x-ray case. Maximum radiation power (top) and electron micro-bunching (middle) with distance through the amplifier, for different $\gamma_m/\gamma_0$. Bottom panel: Longitudinal profiles of radiation (left) and bunching (right) for different $\gamma_m/\gamma_0$: (a) 0~$\%$, (b) 0.04~$\%$, (c) 0.06~$\%$, (d) 0.1~$\%$. Each case is at an equivalent level of micro-bunching. A section of length $\sim l_c$ from a longer bunch is shown; $l_c$ increases with increasing $\gamma_m/\gamma_0$.
   }
   \label{Soft_xray_amplifier}
    \vspace{-6mm}
\end{figure}
The performance of the amplifier stage was optimised by varying the relative amplitude, $\gamma_m/\gamma_0$.
The growth of the radiation power and electron bunching are plotted for a range of electron energy modulation in Fig.~\ref{Soft_xray_amplifier}.
Increasing the energy modulation amplitude decreases the region about the extrema able to lase and the mean amplification rate decreases. However, a pronounced comb in the electron bunching of period $\approx \lambda_m$ is seen to develop.
Since the radiation propagates through the beam, only a relatively small undulation of the radiation power on the scale of $\lambda_m$ is present.
The optimum modulation amplitude was determined to be $\gamma_m/\gamma_0\approx\rho$, with $\gamma_m/\gamma_0$=0.1~\% used for injection into the afterburner.
The extraction point from the amplifier stage was chosen to be 34.1~m, as shown in Fig.~\ref{Soft_xray_amplifier}.
Hence, no increase in the amplifier length from that for normal saturated SASE operation is required.
Both the electron beam and radiation from the amplifier stage propagate into the afterburner (Fig.~\ref{Schematic}).
Each afterburner module has 8 undulator periods followed by a chicane that delays the electron beam by 23 resonant wavelengths, so that the total electron delay per module $s=(8+23)\times\lambda_r=\lambda_m$.
Energy dispersion effects in the chicanes were included, although new chicane designs that reduce dispersive effects may be possible~\cite{D0_chicanes}.
Figure~\ref{ML_AB_output} plots the radiation power and spectrum at different positions in the afterburner.
A pulse train structure develops rapidly as the radiation and bunching combs are regularly re-phased by the chicanes to maintain overlap in the amplifying undulator sections.
The growth within the undulator modules of the afterburner is exponential of rate comparable to that in the amplifier stage with no beam energy modulation.
The growth in the afterburner is also enhanced by the additional bunching caused by the dispersive chicanes~\cite{mlsase}.
After 15 afterburner modules the output consists of a train of $\sim$9~as rms radiation pulses separated by $\sim 124$~as and of $\sim$0.6~GW peak power. The corresponding spectrum is multichromatic with bandwidth envelope increased by $\sim$50 over that of SASE. The pulse train envelope has fluctuations typical of SASE, with phase correlation between individual radiation pulses over a co-operation length.
Each afterburner module consists of an undulator module of length 0.26~m and a chicane of length 0.2~m, giving a total length of 6.9~m (excluding diagnostics etc.) for the 15-module afterburner.

\begin{figure}[t]
   \centering
   \includegraphics[width=0.95\columnwidth]{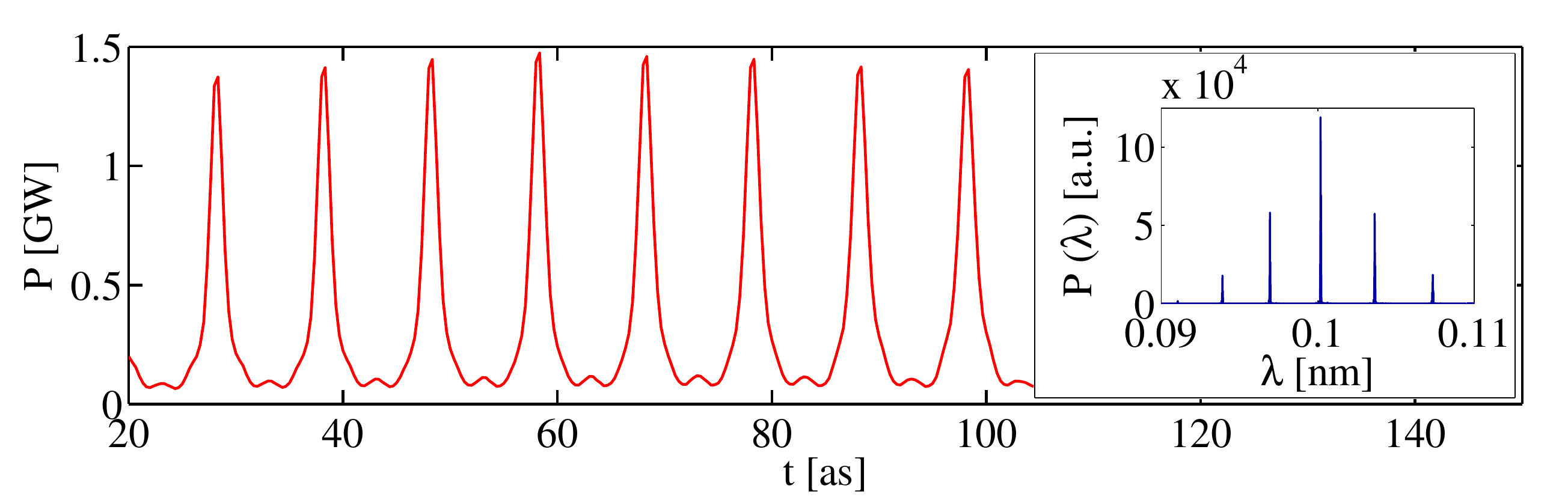}
   \vspace{-2mm}
   \caption{Hard x-ray mode-locked afterburner simulation results: Radiation power profile and spectrum after 40 modules. The duration of an individual pulse is $\sim$700~zs~rms.}
   \label{0p1nm_ML_AB}
\vspace{-6mm}
\end{figure}

A hard x-ray case of resonant FEL wavelength of $\lambda_r$=0.1~nm was also simulated, with the aim of demonstrating shorter pulse generation.
A modulation period of $\lambda_m$=3~nm was used ($\lambda_m$=$30\times\lambda_r$) which may be feasible using HHG sources that are now being developed~\cite{HHG_3nm}. Both the undulator and electron beam parameters used are similar to those of the compact SACLA x-ray FEL facility~\cite{SACLA}, and are detailed in Table~\ref{table1}. A reduced peak current is used, typical for a lower electron bunch charge. This allows a slightly reduced, but still realistic, emittance to be assumed to attain a more compact afterburner stage. As for the soft x-ray case above, the amplifier stage was optimised and a beam energy modulation of $\gamma_m/\gamma_0$=0.06~\% chosen. The amplifier section consists of six 277-period undulator modules (36~m). Each afterburner module consists of an undulator module of 8 periods and a chicane with delay of 22$\times\lambda_r$. The total electron delay per afterburner module is then equal to $\lambda_m$. The total afterburner consists of 40 modules each consisting of an undulator of length 0.144~m and a chicane of length 0.2~m to give 13.8~m in total. Figure~\ref{0p1nm_ML_AB} plots the radiation power and spectrum after 40 undulator-chicane modules.
A pulse train structure of approximately 700~zs rms duration radiation pulses separated by 10~as and of 1.5~GW peak power is generated. The radiation mode separation is determined by the modulation period of 3~nm corresponding to photon energy difference of $\approx$ 412~eV. The final spectrum is multichromatic with bandwidth envelope of the modes increased by a factor $\sim$100 over SASE.

Both examples in this Letter were optimised towards minimising pulse durations using parameters close to those available from current x-ray FEL sources. Using short (8-period) undulator modules in the afterburner, $\sim$5 optical cycles FWHM were attained. However, the time structure could be modified by changing the number of undulator periods, electron delay lengths, and $\lambda_m$, suggesting a development route from present attosecond pulse train experiments~\cite{strobe,apts} to the ultimate capability of the scheme. Amplification in the afterburner was set to occur just before saturation, allowing a short afterburner to attain high contrast ratio of the pulse train over the amplifier radiation. Further development to maximise the peak power and flexibility of the pulse structure could include investigation of saturation effects in the afterburner (e.g. chicane dispersion, radiation diffraction) and their mitigation through e.g.~undulator tapering~\cite{tapering}, optimised phase-shifting~\cite{ratner}, use of low-dispersion~\cite{D0_chicanes} or more compact chicanes. Methods to improve the temporal coherence and stability developed for SASE, may also be applicable.
A potential proof-of-principle experiment would be to use a single-module afterburner~\cite{FEL10 paper}. This is similar to other proposals~\cite{zholents, echo_atto} to attain pulse lengths  $\ll l_{c}$, but at relatively low power, since they operate via coherent spontaneous emission in a single short undulator, rather than exponential amplification.

If the above results are scaled to higher photon energies, e.g to the 50~keV of the proposed x-ray FEL of~\cite{marie}, then pulse durations of 140~zs rms may become feasible. Operation at harmonics of $\lambda_r$ may be another route to shorter pulse durations. Furthermore, if a relativistic counter-propagating target nuclear beam were used with such pulse trains, as discussed in~\cite{nqo}, in addition to the increased doppler-shifted photon energies that the nuclei experience in their boosted frame, the pulse durations may be further reduced towards the timescales of highly ionised heavy elements and nuclear dynamics~\cite{kaplan}.

The mode-locked afterburner is potentially a relatively simple upgrade to existing x-ray FEL facilities. It offers a flexible route towards the generation of discretely multichromatic output under a broad bandwidth envelope, and so offers few-cycle x-ray pulse trains with GW peak-powers in the temporal domain. This would help facilitate the direct study of the temporal evolution of complex correlated electronic behaviour within atoms, and push capability into the regime of electronic-nuclear dynamics and towards that of the nucleus.

\vspace{2mm}
\begin{acknowledgements}
The authors would like to thank Ken Ledingham for helpful discussions.
\vspace{1mm}

This work received support from STFC Memorandum Of Agreement Number 4163192.
\vspace{-3mm}
\end{acknowledgements}

\vspace{0mm}


\begin{thebibliography}{99} 

\bibitem{atto1} P.B. Corkum and F. Krausz, Nature Phys. {\bf 3}, 381 (2007).
\bibitem{atto2} F. Krausz and M. Ivanov, Rev. Mod. Phys. {\bf 81}, 163 (2009).
\bibitem{hhg}A. McPherson \emph{et al.}, JOSA B {\bf 4}, 595 (1987).
\bibitem{strobe} J. Mauritsson \emph{et al.}, Phys. Rev. Lett. \textbf{100}, 073003 (2008).
\bibitem{apts}P. Johnsson \emph{et al.}, Phys. Rev. Lett. {\bf 95}, 013001 (2005),  T. Remetter \emph{et al.}, Nature Phys. {\bf 2}, 323 (2006), K. Varj\'{u} \emph{et al.}, J. Phys. B: At. Mol. Opt. Phys. {\bf 39}, 3983 (2006),  M. Klaiber, K.Z. Hatsagortsyan, C. M\"{u}ller and C.H. Keitel, Optics Lett. {\bf 33}, 411 (2008), K.P. Singh \emph{et al.}, Phys. Rev. Lett. {\bf 104}, 023001 (2010).

\bibitem{neet} S. Kishimoto, Phys. Rev. Lett. {\bf 85}, 1831 (2000).
\bibitem{kaplan} A.E. Kaplan, Lasers in Eng. {\bf 24}, 3 (2012).

\bibitem{FLASH} W. Ackermann, \emph{et al.}, Nature Photon. \textbf{1}, 336-342 (2007).

\bibitem{LCLS} P. Emma \emph{et al.}, Nature Photonics \textbf{4}, 641 - 647 (2010).

\bibitem{SACLA} H. Tanaka, T. Shintake \emph{et al.}, SCSS X-FEL Conceptual Design Report, RIKEN Harima Institute (2005), H. Tanaka \emph{et al.}, Nature Photon.
                 \textbf{6}, 540-544 (2012).

\bibitem{FERMI@elettra} C.J. Bocchetta \emph{et al.} FERMI@Elettra Conceptual Design Report ST/F-TN-07/12 (Sincrotrone Trieste), (2007).

\bibitem{natphoton}
B.W.J. McNeil and N.R. Thompson, Nature Photon. {\bf 4}, 814 (2010).

\bibitem{marie}
B.E. Carlsten \emph{et al.}, Proceedings of 2011 Particle Accelerator Conference, New York, USA, 799-801 (2011).

\bibitem{bnp}R. Bonifacio, L.M. Narducci and C. Pellegrini, Opt. Commun. {\bf 50}, 373 (1984).

\bibitem{bonprl} R. Bonifacio, L. De Salvo, P. Pierini, N. Piovella and C. Pellegrini, Phys. Rev. Lett. \textbf{73}, 70 (1994).

\bibitem{limit}E.L. Saldin, E.A. Schneidmiller and M.V. Yurkov, Opt. Commun. {\bf 212}, 377 (2002).

\bibitem{RIXS} L.J.P. Ament, M. van Veenendaal, T.P. Devereaux, J.P. Hill and J. van den Brink, Rev. Mod. Phys. \textbf{83}, 705, (2011).

\bibitem{nqo}T.J. B\"{u}rvenich, J. Evers and C.H. Keitel, Phys. Rev. Lett. {\bf 96}, 142501 (2006).

\bibitem{bmp}R. Bonifacio, B.W.J. McNeil and P. Pierini, Phys. Rev. A {\bf 40}, 4467 (1989).

\bibitem{saldin}E.L. Saldin, E.A. Schneidmiller and M.V. Yurkov, Phys. Rev. ST Accel. Beams {\bf 9}, 050702 (2006).

\bibitem{zholentsNJP}A.A. Zholents and M.S. Zolotorev, New Journal of Physics {\bf 10}, 025005 (2008).

\bibitem{emma}P. Emma, Z. Huang and M. Borland, Proc. 26\textsuperscript{th} Int. FEL Conf., Trieste, 333 (2004).

\bibitem{ml_esase} D. Xiang, Y. Ding and T. Raubenheimer and J. Wu, Phys. Rev. ST Accel. Beams {\bf 15}, 050707 (2012).

\bibitem{shortpulselcls}Y. Ding \emph{et al.}, Phys. Rev. Lett. \textbf{102} 254801, (2009).

\bibitem{mlsase} N.R. Thompson and B.W.J. McNeil, Phys. Rev. Lett. {\bf 100}, 203901 (2008).

\bibitem{s2e_mlfel}
D.J. Dunning, B.W.J. McNeil, N.R. Thompson and P.H. Williams, Physics of Plasmas, \textbf{18}, 073104, (2011).

\bibitem{esase} A.A. Zholents, Phys. Rev. ST Accel. Beams \textbf{8}, 040701 (2005).

\bibitem{njp_ml_curmod}
E. Kur, D.J. Dunning, B.W.J. McNeil, J. Wurtele and A.A. Zholents, New J. Phys. \textbf{13} 063012 (2011).

\bibitem{linear}B.W.J. McNeil, `Linear theory of an FEL with an energy modulated electron beam', (unpublished).

\bibitem{genesis}
S. Reiche, Nucl. Inst. Meth. Phys. Res. A, \textbf{429}, 243 (1999).

\bibitem{CDR}
J.P. Marangos \emph{et al.}, New Light Source Project: Conceptual Design Report, Science \& Technology Facilities Council (2010).

\bibitem{D0_chicanes} J.K. Jones, J.A. Clarke and N.R.Thompson, in Proceedings of IPAC 2012, New Orleans, USA, 1759 (2012).

\bibitem{HHG_3nm} T. Popmintchev \emph{et al.}, Science \textbf{336}, 1287 (2012).

\bibitem{tapering}
N.M. Kroll, P.L. Morton and M.N. Rosenbluth, IEEE J. Quant. Electron. \textbf{17}, 1436 (1981).

\bibitem{ratner}D. Ratner, A. Chao and Z. Huang, Proc. 29\textsuperscript{th} Int. FEL Conf., Novosibirsk, 71 (2007).

\bibitem{FEL10 paper} D.J. Dunning, N.R. Thompson, B.W.J. McNeil, Proc. 32\textsuperscript{nd} Int. FEL Conf., Malm\"{o}, Sweden, 636 (2010).

\bibitem{zholents} A.A. Zholents, W.M. Fawley, Phys. Rev, Lett. \textbf{92}, 224801 (2004).

\bibitem{echo_atto} D. Xiang, Z. Huang and G. Stupakov, Phys. Rev. ST Accel. Beams {\bf 12}, 060701 (2009).

\end{thebibliography}
\end{document}